\documentclass{vgtc}                          

\ifpdf
  \pdfoutput=1\relax                   
  \usepackage{graphicx}                
  \DeclareGraphicsExtensions{.pdf,.png,.jpg,.jpeg} 
\else
  \usepackage{graphicx}                
  \DeclareGraphicsExtensions{.eps}     
\fi%

\graphicspath{{figures/}{pictures/}{images/}{./}} 

\usepackage{microtype}                 
\PassOptionsToPackage{warn}{textcomp}  
\usepackage{textcomp}                  
\usepackage{mathptmx}                  
\usepackage{times}                     
\usepackage{cite}                      
\usepackage{tabu}                      
\usepackage{booktabs}                  
\usepackage{multirow}
\usepackage{enumitem}

\onlineid{0}

\vgtccategory{Research}

\vgtcinsertpkg

\title{Evaluation of cinematic volume rendering open-source and commercial solutions for the exploration of congenital heart data}

\author{Irum Baseer\thanks{e-mail: irum.baseer@upf.edu}\\ %
     \parbox{1.4in}{\scriptsize \centering  Universitat Pompeu Fabra} %
 \and Israel Valverde\thanks{e-mail: ivalverde-ibis@us.es }\\ %
      \parbox{1.4in}{\scriptsize \centering  Institute of Biomedicine of Seville} %
 \and Abdel H. Moustafa\thanks{e-mail:amoustafa@santpau.cat }\\ %
     \parbox{1.4in}{\scriptsize \centering Hospital de la Santa Creu i Sant Pau} %
\and Josep Blat\thanks{e-mail:josep.blat@upf.edu}\\ %
     \parbox{1.4in}{\scriptsize \centering Universitat Pompeu Fabra} %
 \and Oscar Camara\thanks{e-mail: oscar.camara@upf.edu}\\ %
      \parbox{1.4in}{\scriptsize \centering Universitat Pompeu Fabra}
}

\abstract{Detailed anatomical information is essential to optimize medical decisions for surgical and pre-operative planning in patients with congenital heart disease. The visualization techniques commonly used in clinical routine for the exploration of complex cardiac data are based on multi-planar reformations, maximum intensity projection, and volume rendering, which rely on basic lighting models prone to image distortion. On the other hand, cinematic rendering (CR), a three-dimensional visualization technique based on physically-based rendering methods, can create volumetric images with high fidelity. However, there are a lot of parameters involved in CR that affect the visualization results, thus being dependent on the user's experience and requiring detailed evaluation protocols to compare available solutions. In this study, we have analyzed the impact of the most relevant parameters in a CR pipeline developed in the open-source version of the MeVisLab framework for the visualization of the heart anatomy of three congenital patients and two adults from CT images. The resulting visualizations were compared to a commercial tool used in the clinics with a questionnaire filled in by clinical users, providing similar definitions of structures, depth perception, texture appearance, realism, and diagnostic ability.%
} 

\CCScatlist{
  \CCScatTwelve{Cinematic rendering}{open-source}{commercial tool}{congenital heart data}{}
}

\begin{document}


\firstsection{Introduction}

\maketitle

Congenital heart disease (CHD) is one of the most frequently diagnosed defects afflicting approximately 0.8\% to 1.2\% of live births worldwide \cite{bouma2017changing}. As the cardiovascular morphology varies greatly between individual patients, it is important for clinicians to have a comprehensive understanding of the spatial relationship between the cardiac structures, in order to make optimal medical decisions.  As a result, there is a growing number of computer-aided software tools available to assist radiologists in this process. One of the widely used methods is volume rendering, which was found to represent human structures in an artificial way, also being prone to image distortions \cite{eid2017cinematic} due to their reliance on basic lighting models. On the other hand, cinematic rendering (CR) is a novel post-processing tool \cite{fellner2016introducing} that renders the volumetric medical data using physically-based advanced lighting models \cite{kroes2012exposure}. CR resembles the casting of billions of light rays from all possible directions to create volumetric images with a remarkable level of realism \cite{dappa2016cinematic}. 
\begin{figure*}
    \centering
    \tabcolsep=0.03cm
    \renewcommand{\arraystretch}{0.13}
    \begin{tabular}{ccccc}
    \includegraphics[width=0.17\textwidth,height=0.17\textwidth]{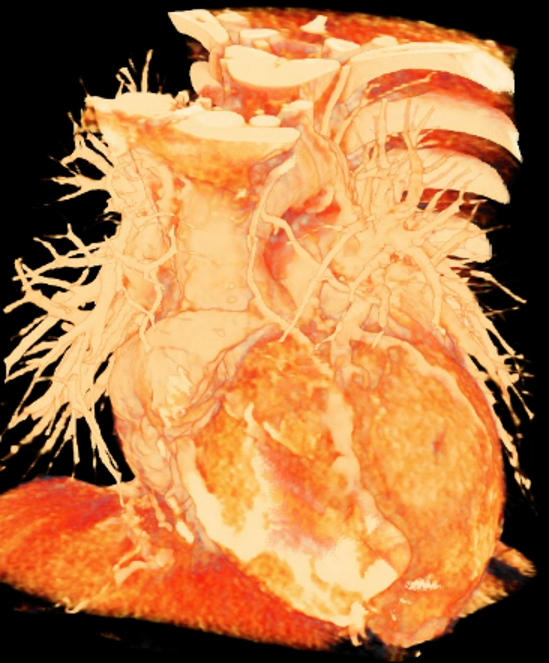}&   \includegraphics[width=0.17\textwidth,height=0.17\textwidth]{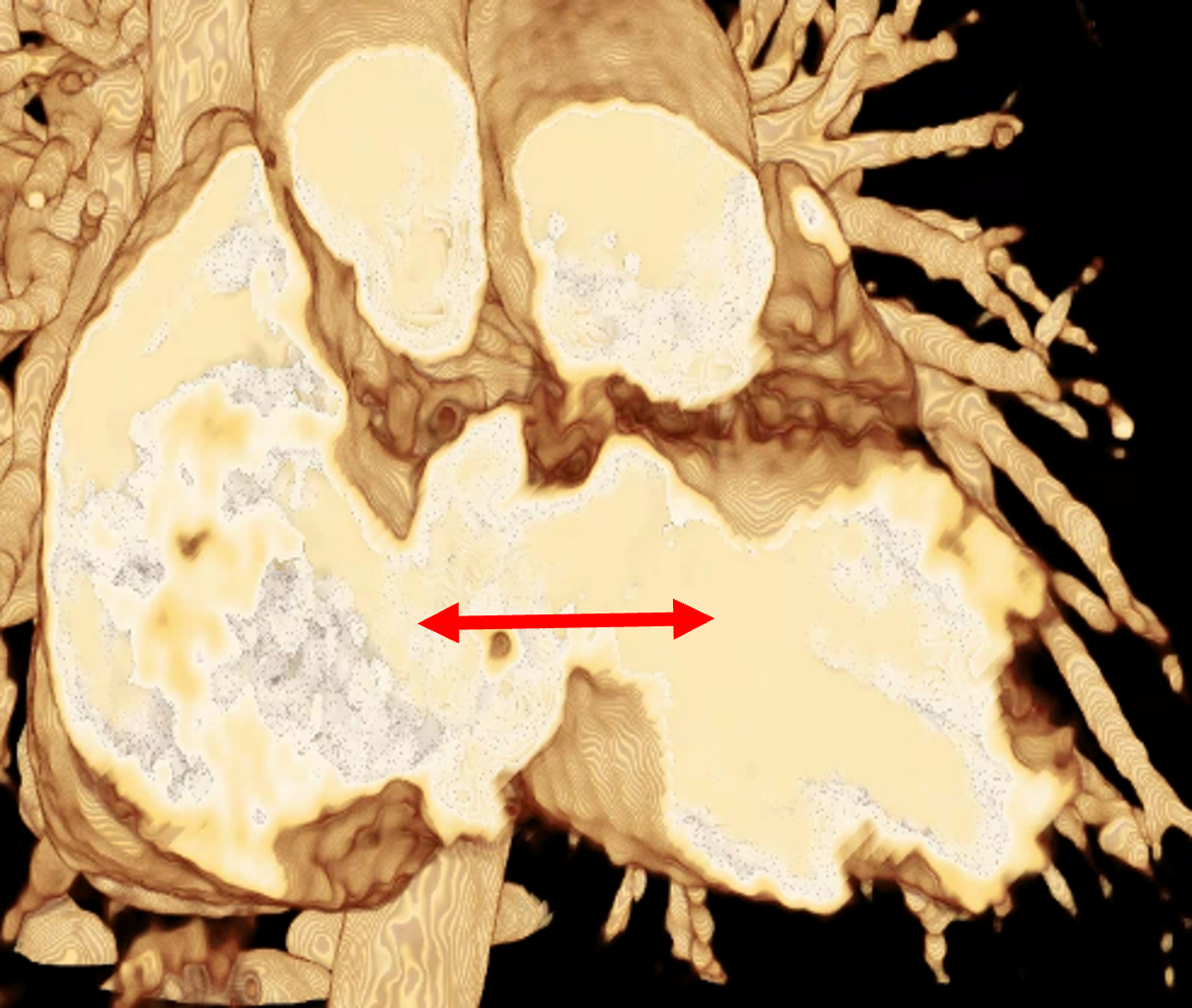}&
    \includegraphics[width=0.17\textwidth,height=0.17\textwidth]{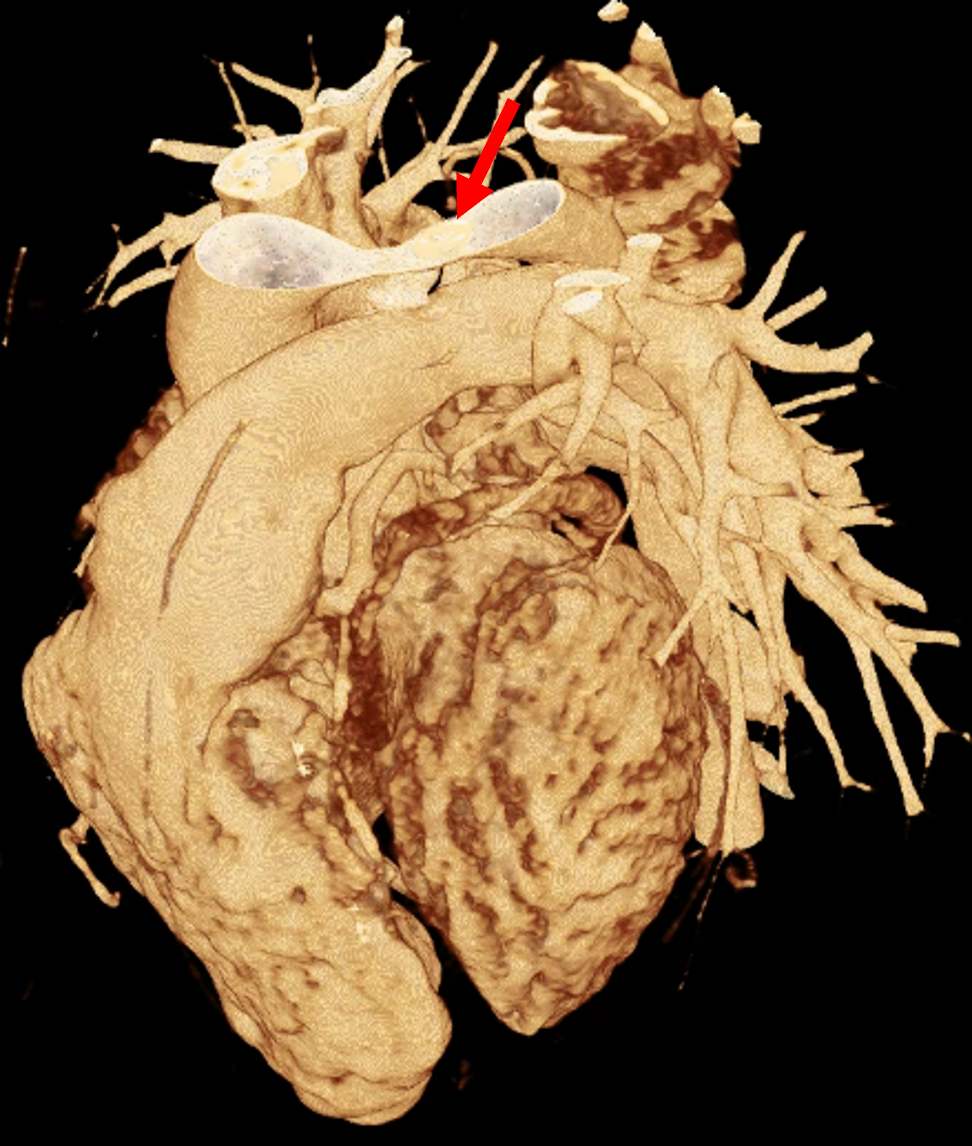}&  \includegraphics[width=0.17\textwidth,height=0.17\textwidth]{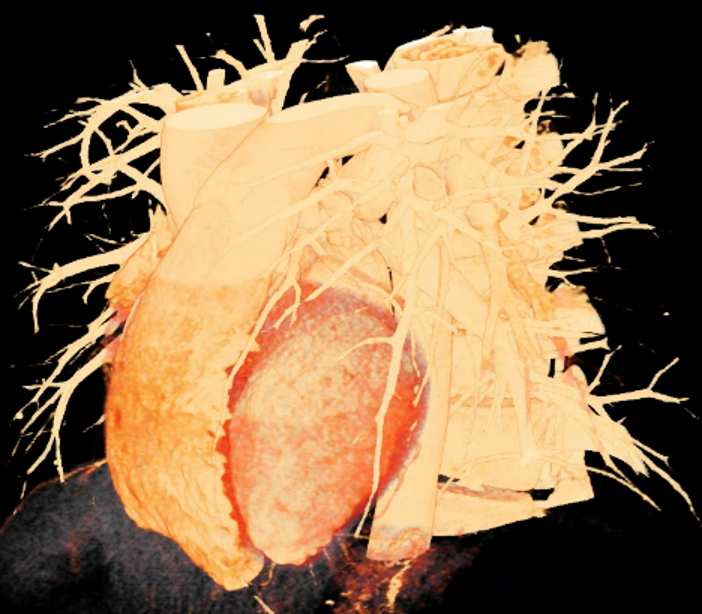}&
    \includegraphics[width=0.17\textwidth,height=0.17\textwidth]{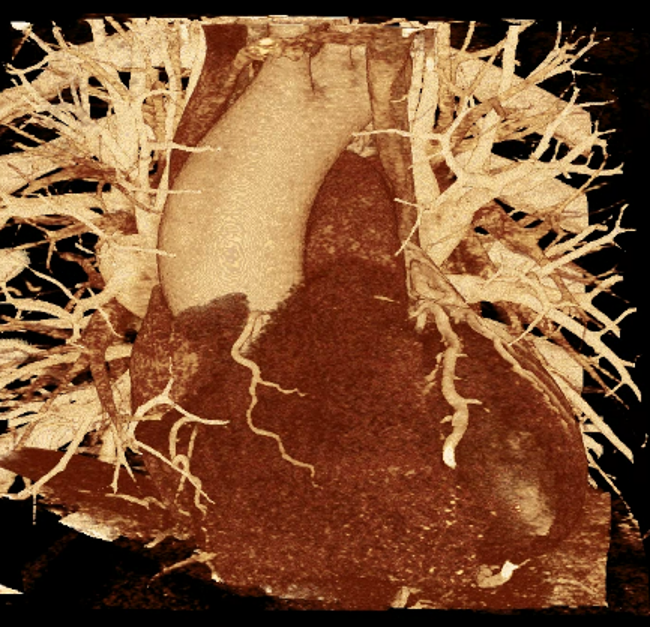}\\
    \includegraphics[width=0.17\textwidth,height=0.17\textwidth]{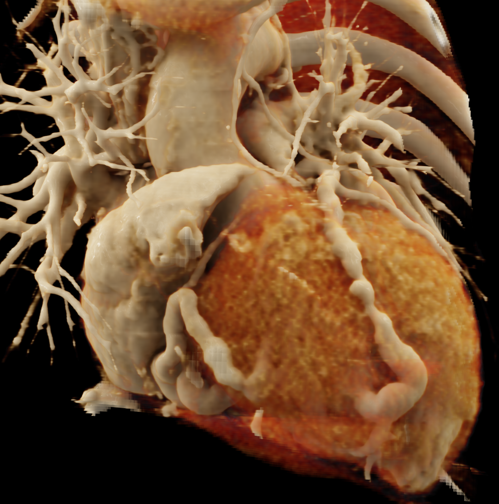}&   \includegraphics[width=0.17\textwidth,height=0.17\textwidth]{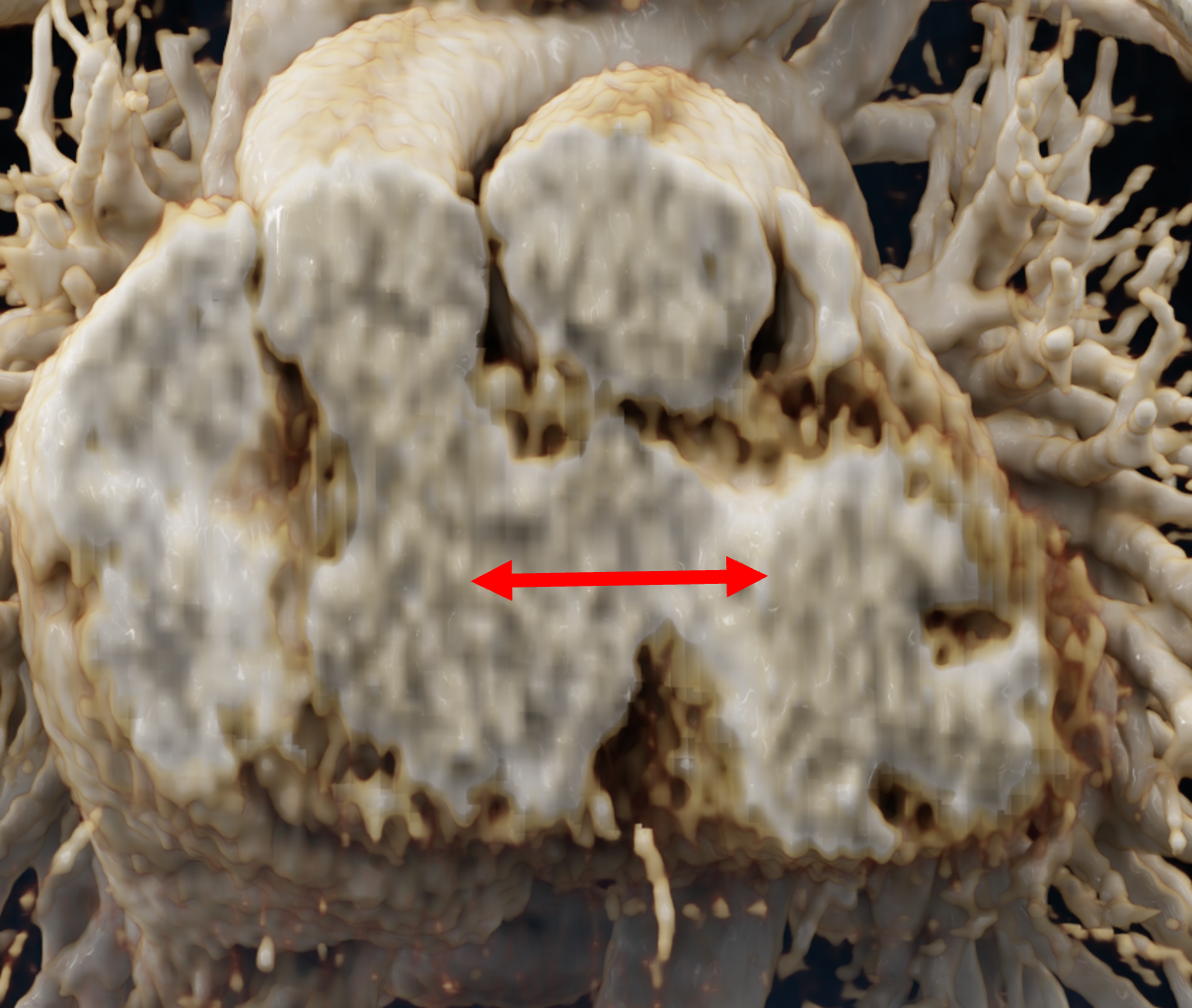}&
    \includegraphics[width=0.17\textwidth,height=0.17\textwidth]{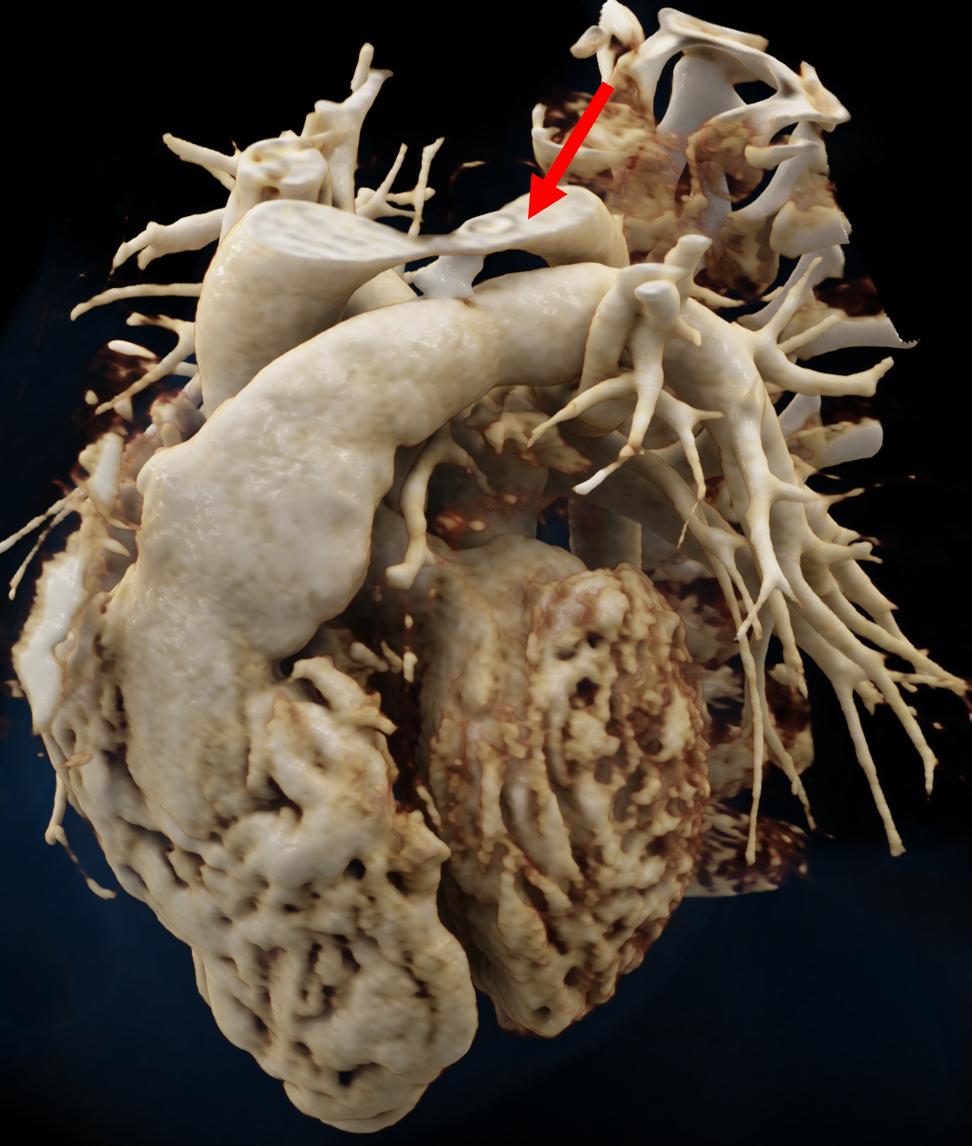}&  \includegraphics[width=0.17\textwidth,height=0.17\textwidth]{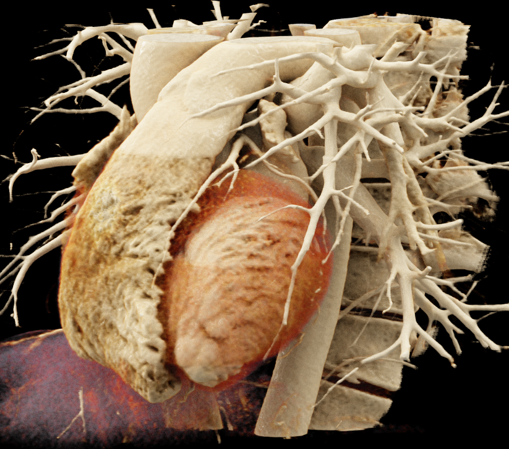}&
    \includegraphics[width=0.17\textwidth,height=0.17\textwidth]{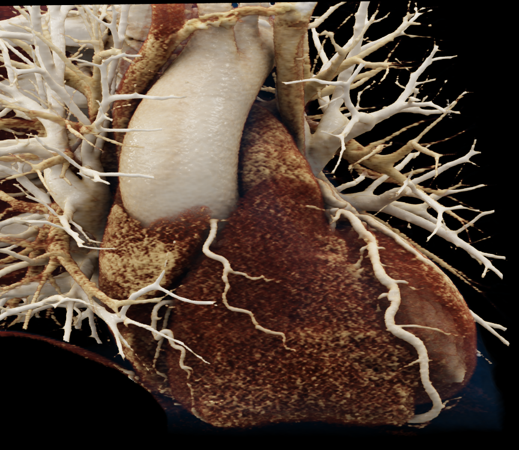}\\
    \includegraphics[width=0.17\textwidth,height=0.17\textwidth]{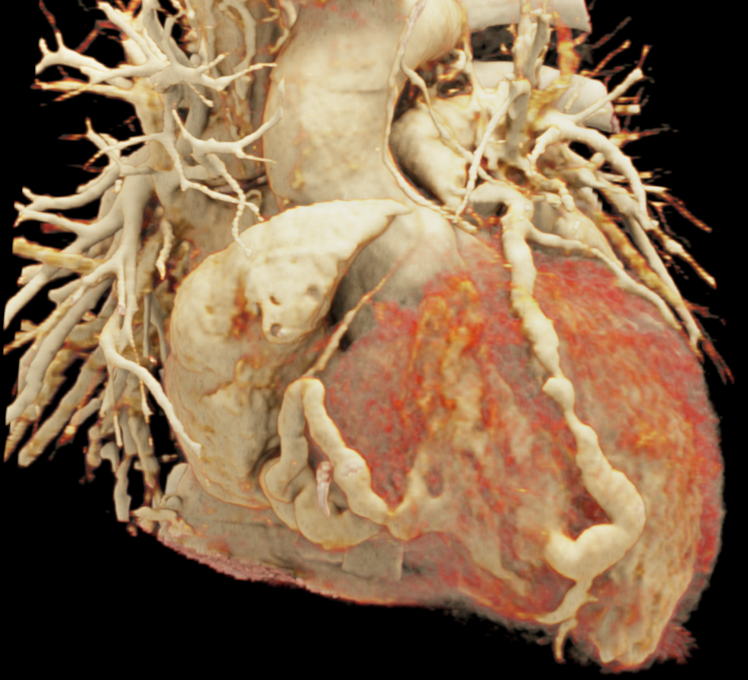}&
    \includegraphics[width=0.17\textwidth,height=0.17\textwidth]{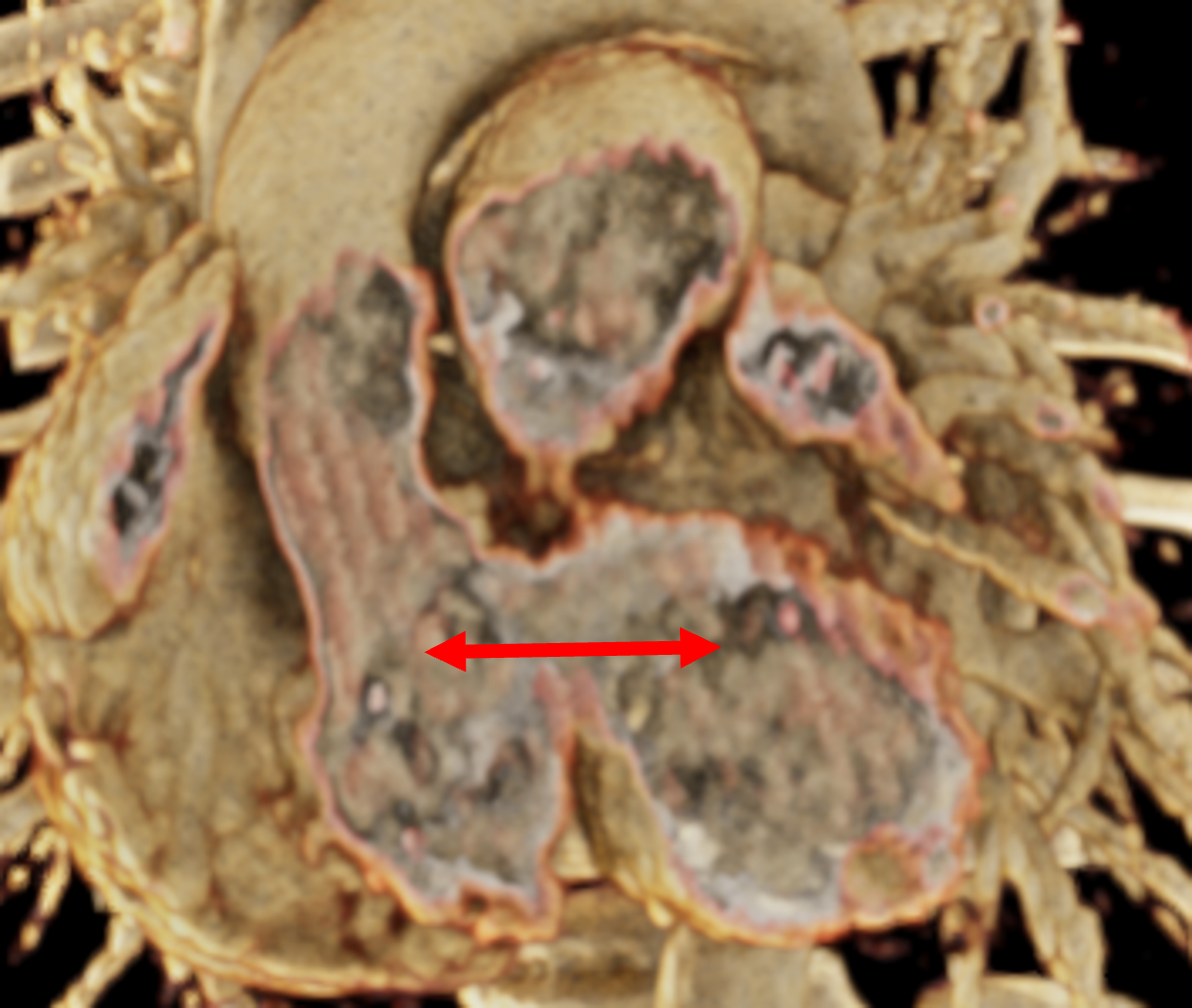}&
    \includegraphics[width=0.17\textwidth,height=0.17\textwidth]{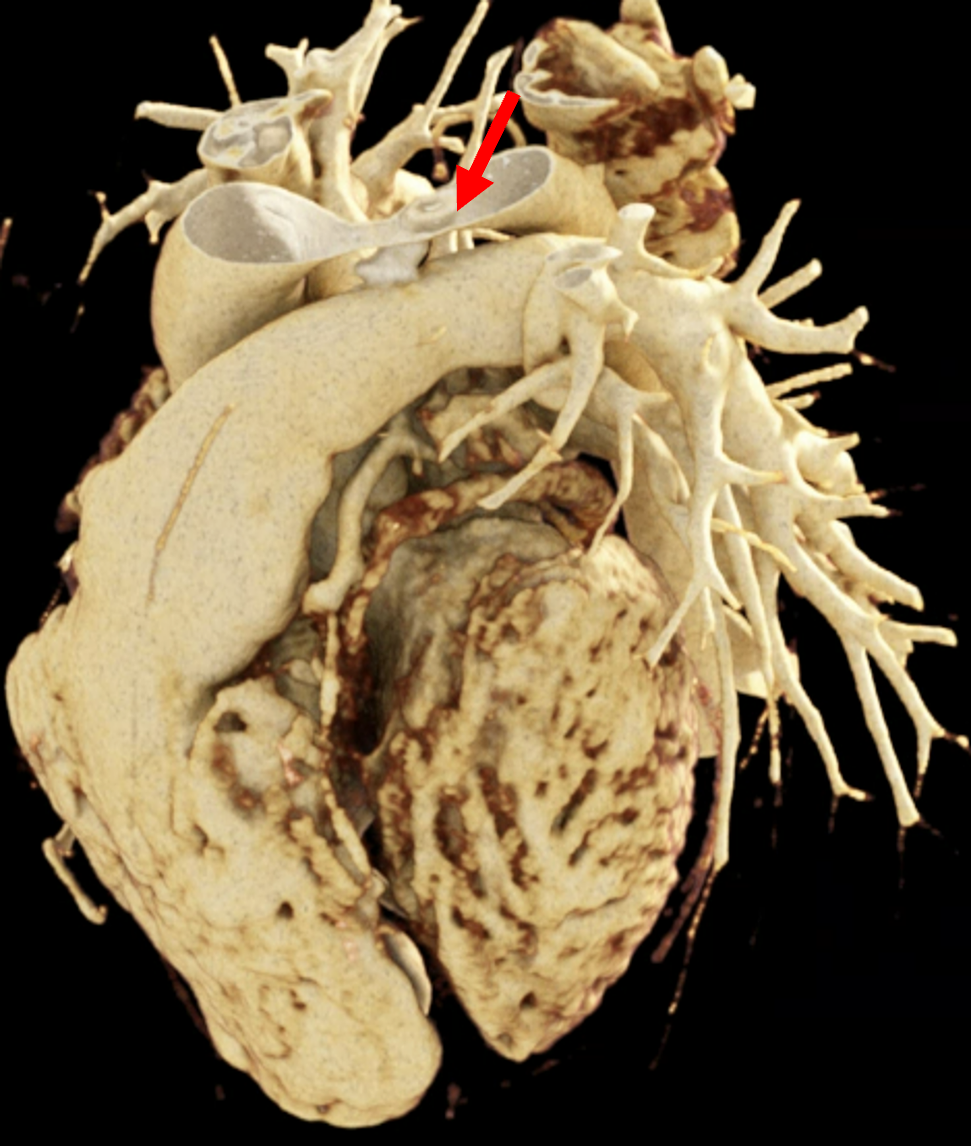}&
    \includegraphics[width=0.17\textwidth,height=0.17\textwidth]{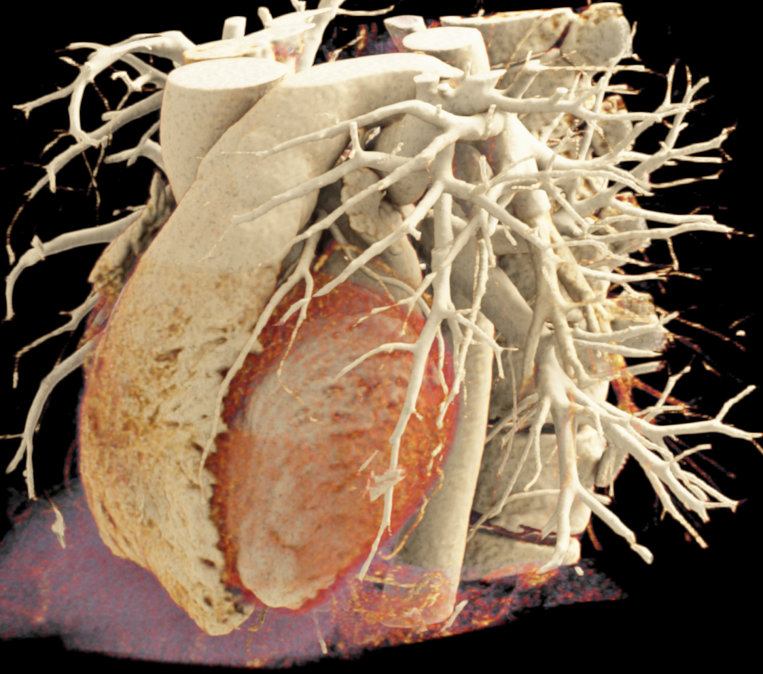}&
    \includegraphics[width=0.17\textwidth,height=0.17\textwidth]{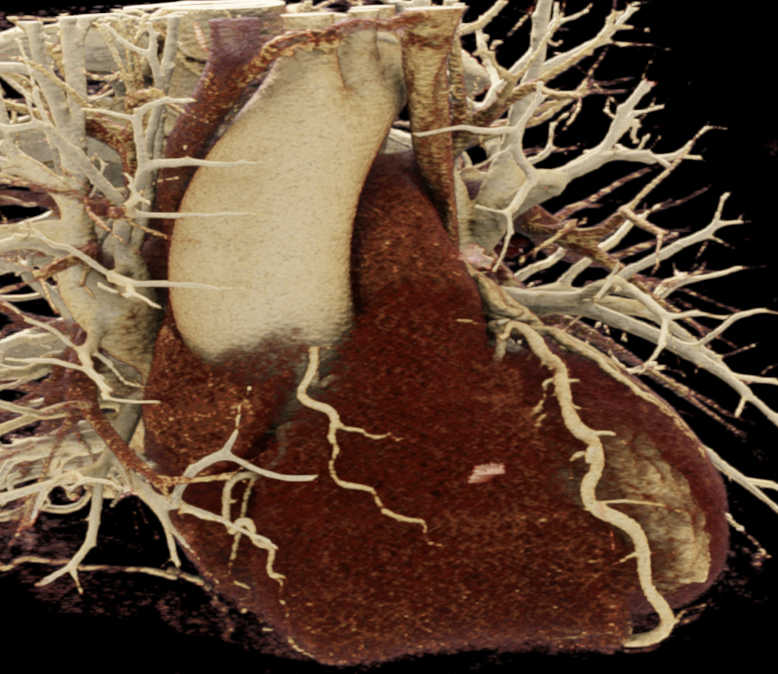}
         
    \end{tabular}
    
    \caption{Comparison of volume rendering (VR) and cinematic rendering (CR) visualizations for congenital heart disease patients (Cases 1-3) and normal adults (Cases 4-5). Top row: VR. Middle row: CR from commercial solution. Bottom row: CR from open-source solution.}
    \vspace{-1.5em}
    \label{fig:Figure_1}
\end{figure*} 

Benefits of cinematic vs volume rendering are increasingly being reported in medical applications, such as for faster comprehension of anatomy \cite{elshafei2019comparison} and for pre-operative planning\cite{wollschlaeger2020ct}. Additionally, studies have shown that more accurate visualization of medical data benefits imaging tasks including delineating complex congenital heart pathologies\cite{roschl2019initial}. Apart from its clinical utility, CR also has the potential to be useful in patient communication \cite{ebert2017forensic} and education \cite{binder2019leveraging}.\par
The most widely used commercial implementation of CR\cite{comaniciu2016shaping} is offered by Siemens Healthineers as part of their syngo.via platform\footnote{https://www.siemenshealthineers.com/digital-health-solutions/cinematic-rendering}. Several other license-limited solutions include Global Illumination Vitrea by Canon Medical Informatics\footnote{https://www.vitalimages.com/global-illumination/} and MeVisLab\footnote{https://www.mevislab.de/download/} are also available. Vitrea offers global illumination methods for rendering volumetric data in a photo-realistic manner, while MeVisLab offers a path tracer module, which is a significantly enhanced version of the ExposureRender \cite{kroes2012exposure} framework by Thomas Kroes.
For instance, MeVisLab has recently been used for post-surgical assessment in oncologic head and neck reconstructive surgery comparing path tracing and volume rendering techniques \cite{jimaging9020024}. While these vendor-provided solutions often have high rendering capabilities and are utilized in advanced healthcare centers, their cost can be prohibitive for smaller institutions or individual researchers. However, there are some open-source alternatives. Voreen\footnote{https://www.uni-muenster.de/Voreen} and Inviwo\footnote{http://www.inviwo.org} offer better volumetric rendering capabilities by implementing ray casting with global illumination. Yet, neither of these applications utilizes volumetric path tracing or equivalent state-of-the-art volumetric rendering techniques. Another open-source and freely available solution supporting CR in web browsers is VolView \cite{xu2022interactive}. These solutions are typically free to use, customizable to meet specific needs, and can be used by anyone with an internet connection, regardless of their location or financial resources. While these solutions are affordable and flexible, there is a lack of research comparing different solutions in cardiac applications, to identify the strengths and weaknesses of open-source software tools as compared to commercial solutions.\par
In this study, we utilized a free version of MeVisLab to design a pipeline for cinematically rendering a CHD dataset.  
We conducted a detailed evaluation of several critical parameters to enhance the shape and depth perception of the heart anatomy. Furthermore, we assessed the performance of the developed open-source rendering pipeline by comparing it with a commercial solution available in clinics. This evaluation was conducted using a questionnaire filled out by cardiology experts.

\section{Materials and methods}

\subsection{Patient cases and reconstructions}
We included clinical data of 3 congenital heart disease patients who underwent CT imaging for diagnosis or treatment, along with CT data of two normal adult hearts. A brief summary of patient data is provided in Table \ref{tab:patient-cases}. For creating CR visualizations from CT data, anonymized DICOM reconstructions were transferred to both used environments: MeVisLab and a workstation with prototype commercial software (syngo.via cinematic VRT, Siemens Healthineers). Each case was carefully displayed using the same zooming and rotation features in both software; the cutting tool was applied to eliminate any bones and remaining lines that could obscure the view. To ensure comparability, the generated reconstructions were captured using the same angle of view, color, and opacity settings (see Figure \ref{fig:Figure_1}).
\begin{table}[h]
\caption{Patient cases and reconstructions. M/F: male/female.}  \label{tab:patient-cases}
 \centering
\resizebox{8.45 cm}{!}{
\tabcolsep=0.1cm
    \begin{tabular}{|c|c|c|c|c|c|}
    
       \hline 
 \textbf{Case} & \textbf{M/F} & \textbf{Age}& \textbf{Condition} & \textbf{Manufacturer} & \textbf{Voxel spacing}\\  \hline
1 & F & 3 years & Pulmonary atresia & GE Medical  & $0.273\times0.273\times0.625$\\ \hline

2 & M & 4 days & Ventricular septal defect & Canon Medical & $0.163\times0.163\times0.250$\\ \hline

3 & M & 2 years & Occluded arterial duct & Siemens & $0.246\times0.246\times0.400$\\ \hline

4 & F & 46 years & Normal heart & GE Medical & $0.559\times0.559\times0.625$\\ \hline

5 & M & 63 years & Normal heart & Philips & $0.576\times0.576\times0.329$ \\ \hline 
\end{tabular}
}
\vspace{-1.7em}
\end{table}

\subsection{Cinematic rendering pipeline and sensitivity analysis}
To build the cinematic rendering pipeline, the MeVisLab software (version 3.5.0) was installed on a personal computer (AMD FX (tm), eight-core processor, 3.50GHz, 32 GB RAM, 64-bits operating system). The pipeline involved a visual programming approach, combining various modules to load imaging data, perform pre-processing, design transfer functions, set material and lighting properties, and render the volume followed by post-processing. These steps are briefly explained below.


\textbf{Data loading and pre-processing.} The first step of the rendering involved loading patient data, usually in the form of a series of slices. MeVisLab's DirectDicomImport module was used to directly import the image files. To reduce noise in the acquired data, the GaussSmoothing module in MeVisLab was applied.

\textbf{Transfer function.} After pre-processing, the next step involved designing a transfer function. A transfer function maps voxel values to visual properties like color and opacity, enabling the distinction of anatomical structures in the image. The SoLUTEditor module allowed interactive editing of RGBA lookup tables to design transfer functions. We generated a range of preset transfer functions, saving them in CSV format for easy loading via a Python script. The module's window level and width option allowed for setting a color range for displaying specific anatomical structures, similar to conventional CT reconstructions.

\textbf{Shading and lighting.} Once the data was mapped to the transfer function, lighting and shading were applied to the volume using a physically-based rendering (PBR) workflow. PBR has two principal workflows\cite{biomed}: metal/roughness and specular/glossiness. In PBR, shading is achieved using various Bidirectional Reflectance Distribution Function (BRDF), which are mathematical models that describe how light reflects off surfaces based on their physical properties. The SoPathTracerMaterial module in MeVisLab provides a range of different materials from which Material\textunderscore Microfacet is based on the specular/glossiness workflow of PBR, whereas Material\textunderscore Principled is based on the metal/roughness workflow. We used 'Material\textunderscore Principled', a physically-based material whose parameters are based on the Disney BRDF model\cite{burley2012physically}. We adjusted parameters like base color, metallic, roughness, and specular properties to achieve realistic material appearances.
For lighting, the pipeline included the SoPathTracerAreaLight and SoPathTracerBackgroundLight modules to simulate realistic lighting effects in the 3D scene. The intensity, color, and position of both lights were carefully adjusted to create the desired lighting effect.

\begin{figure}[h]
    \centering
    \tabcolsep=0.03cm
    \begin{tabular}{ccc}
        \includegraphics[width=2.5cm,height=2.5cm]{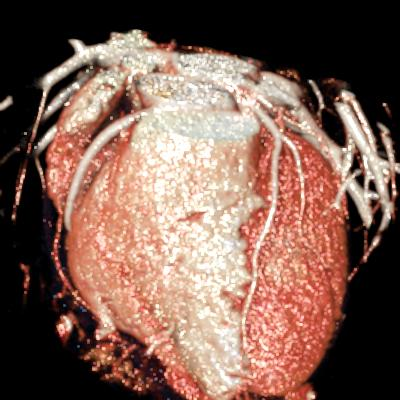}&\includegraphics[width=2.5cm,height=2.5cm]{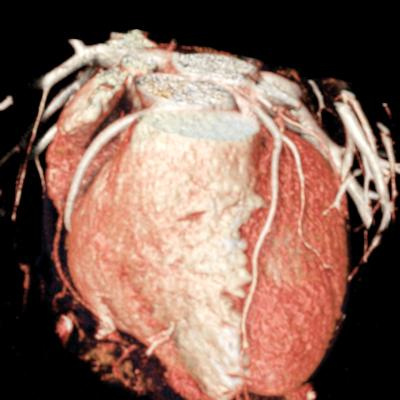}&\includegraphics[width=2.5cm,height=2.5cm]{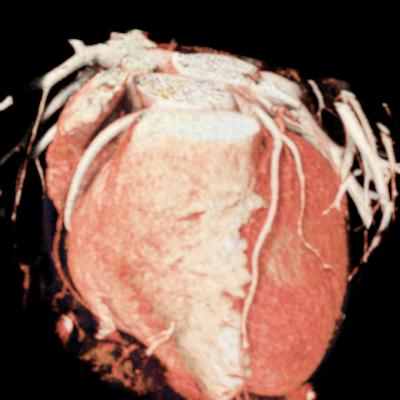} \\
        (a)&(b)&(c)
      \end{tabular}
      \caption{Effects of different values of roughness (r): (a) r = 0, results in a completely shiny and metallic surface (b) r = 0.5, introduces some subtle variations in the surface texture, resulting in a more natural appearance. and (c) r = 1, shows a highly rough surface, leading to a matte surface from left to right, keeping metallic and specular equal to 0.5.}
      \vspace{-1.5em}
    \label{fig:material}
    
\end{figure}
\begin{figure}[ht!]
    \centering
    \tabcolsep=0.03cm
    \begin{tabular}{cc}
       \includegraphics[width=3.5cm,height=2.5cm]{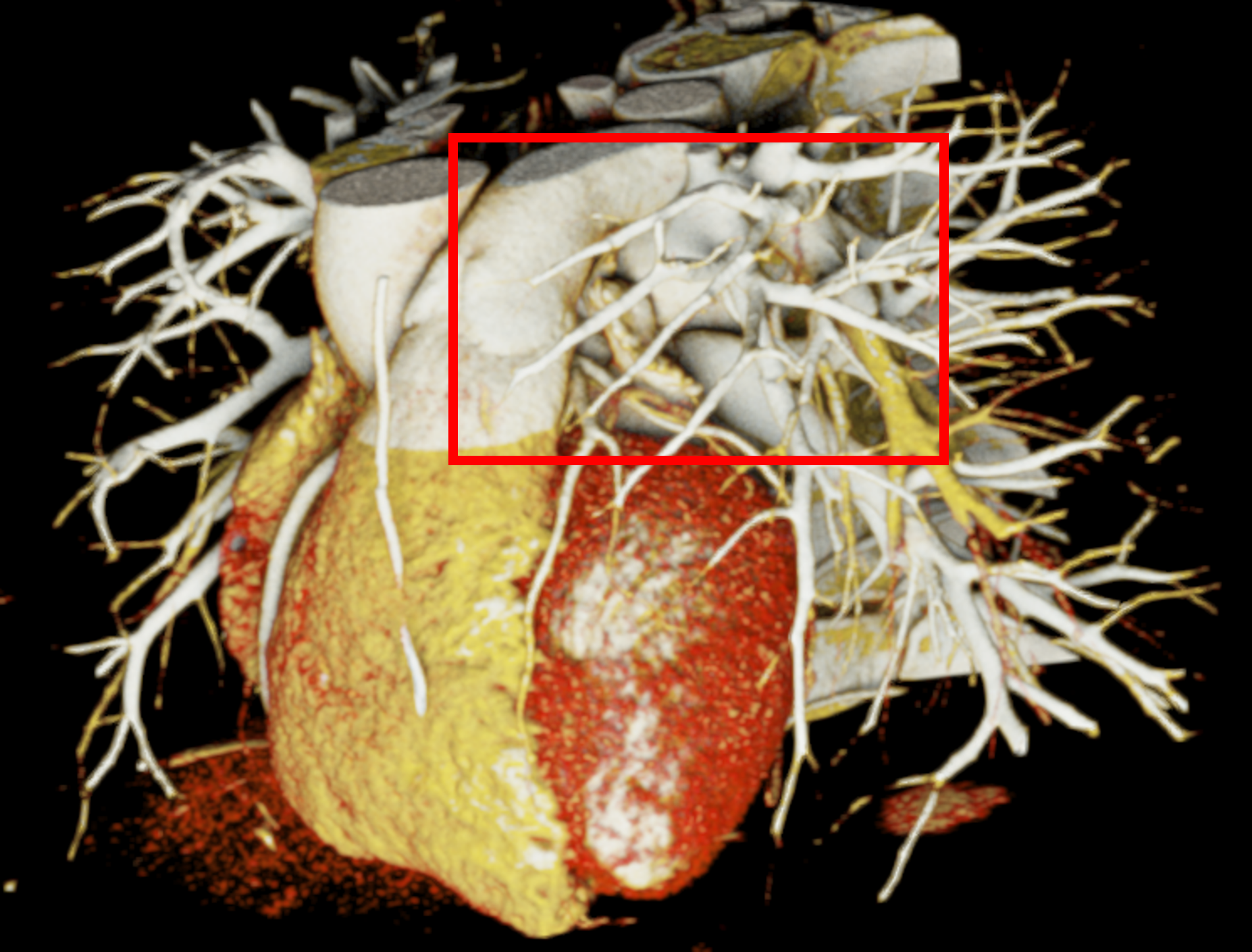}&
       \includegraphics[width=3.5cm,height=2.5cm]{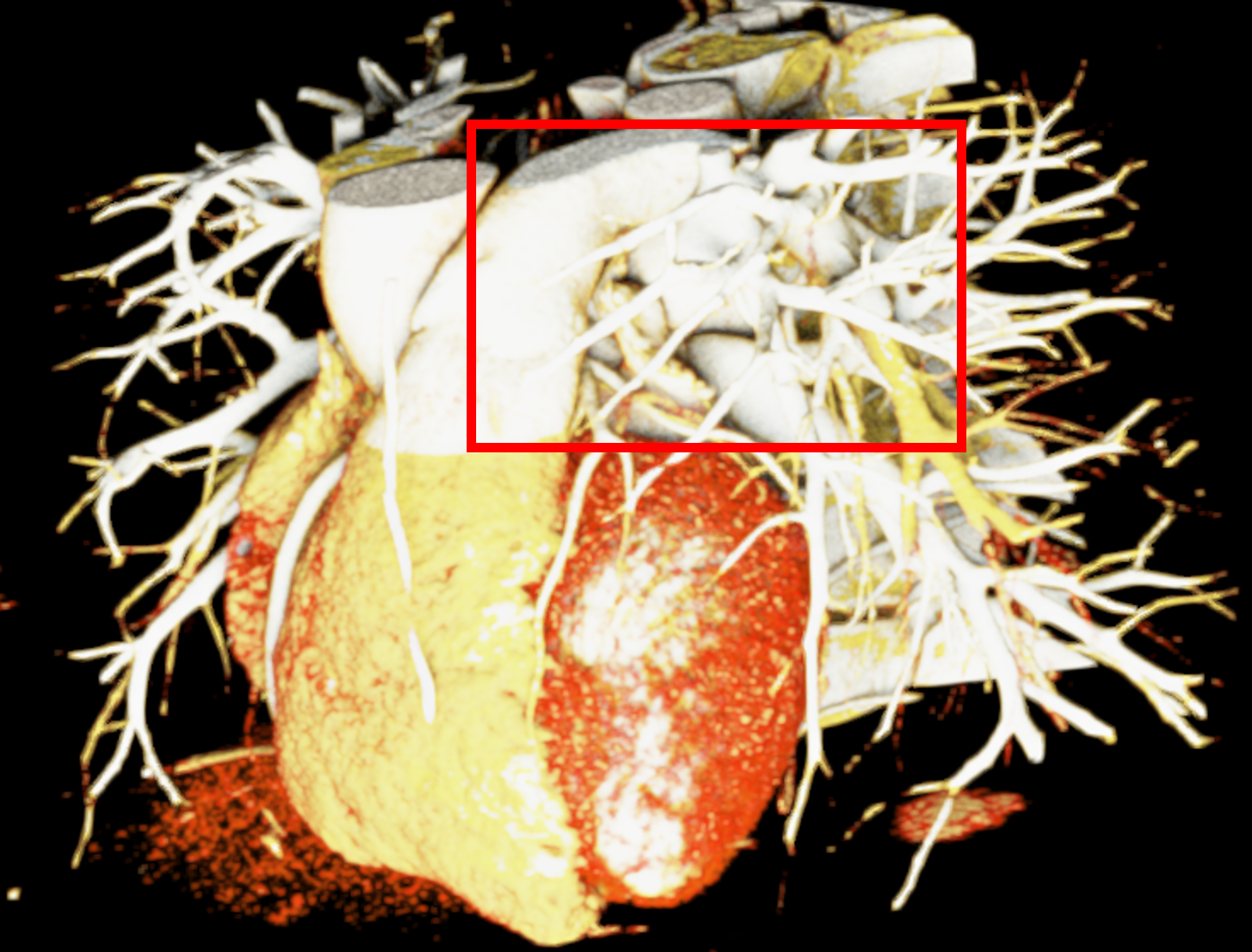}\\
        (a)&(b)
      \end{tabular}
      \caption{Effects of single and multiple lights at the same position. Volume is illuminated with (a) single light, and (b) two lights.}
      \vspace{-1.5em}
    \label{fig:single_two}
\end{figure}
\begin{figure}[ht!]
    \centering
    \tabcolsep=0.03cm
    \begin{tabular}{ccc}
        \includegraphics[width=2.5cm,height=2.5cm]{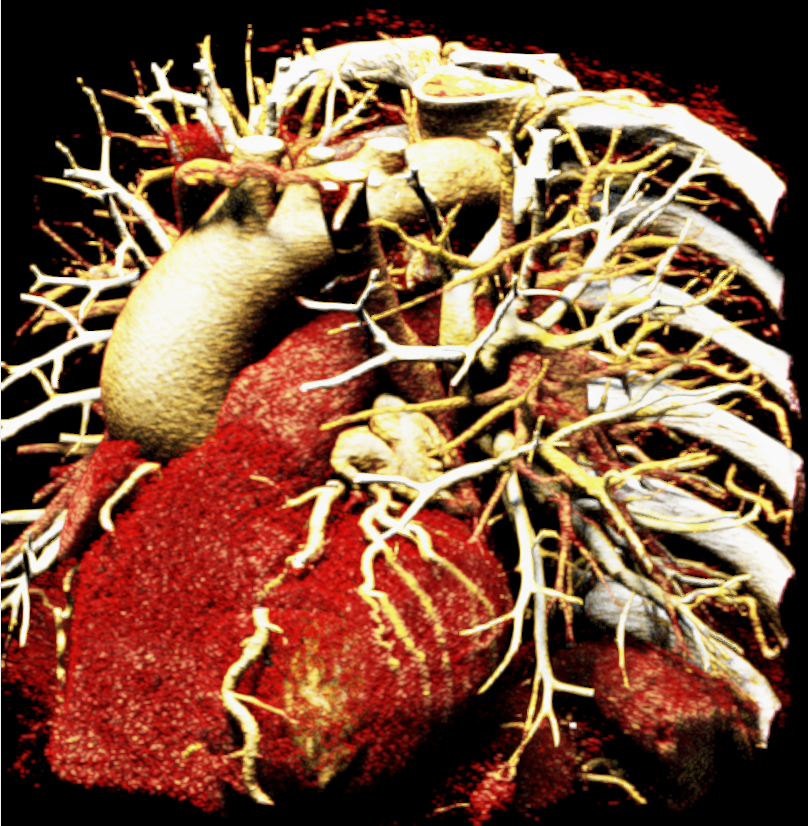}&\includegraphics[width=2.5cm,height=2.5cm]{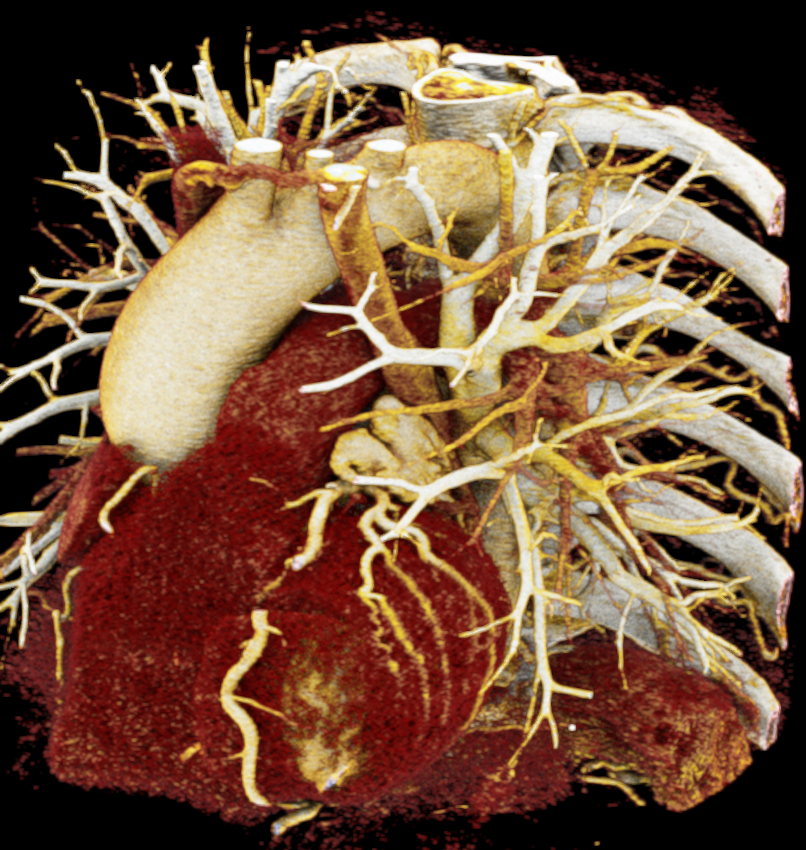}&\includegraphics[width=2.5cm,height=2.5cm]{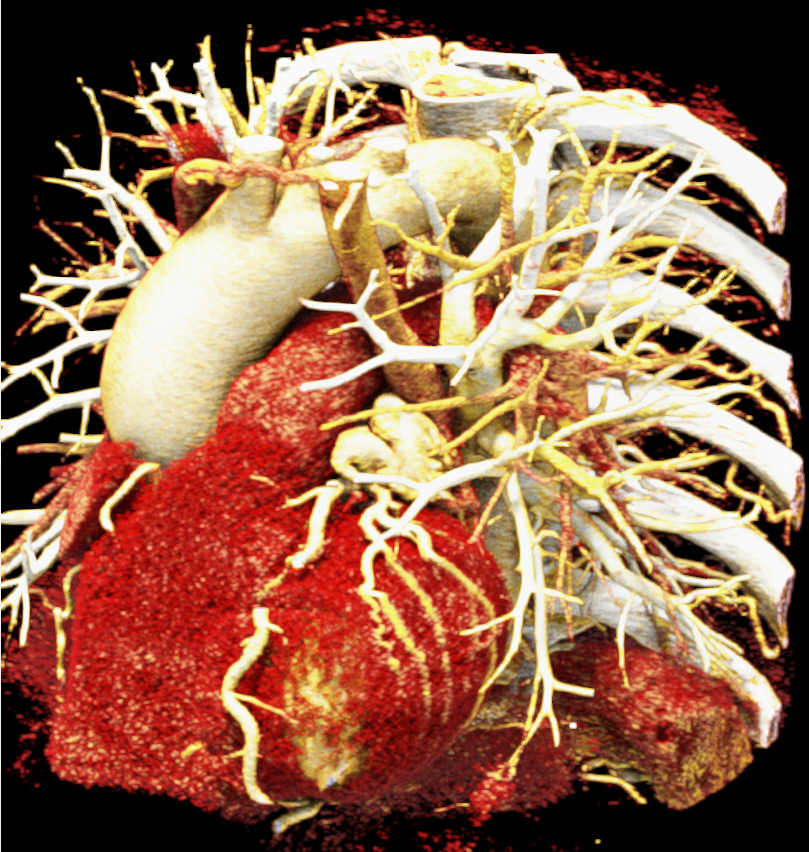}\\
        (a)&(b)&(c)
      \end{tabular}
      \caption{Effects of area and background lighting. Volume is illuminated with: (a) area lighting positioned on the top right; (b) background white lighting; and (c) combination of area and background lighting.}

    \label{fig:directindirect}
\end{figure}

In order to determine which parameters should be adjusted and which values achieved the most realistic appearance, a sensitivity analysis was performed. Specifically, the roughness parameter was tested with constant metallic and specular values of 0.5, producing three images with different roughness levels, as shown in Figure \ref{fig:material}. The impact of lighting on the final image was also examined by varying the number and position of lights. For instance, adding two light sources in the same position tends to create blurry reflections (overexposure) as compared to a single light source, which focuses better on the details of the image as can be seen in Figure \ref{fig:single_two}. Moreover, adjusting the position of light sources improved the visualization of specific regions, including shadows and depth. We also tested the effects of area and background lighting, where area lighting is positioned on the top right and the background light source is positioned behind the objects being lit, providing overall illumination of the scene. As illustrated in Figure \ref{fig:directindirect}, the use of only area lighting tended to overexpose certain areas and obscure details. On the other hand, the use of background white lighting alone resulted in a more diffused and natural-looking, while combining area and background lighting produced dynamic effects, highlighting specific features of the volume with area light, and creating a sense of depth and dimensionality using the background light.

\begin{figure}[ht!]
    \centering
    \tabcolsep=0.02cm
    \renewcommand{\arraystretch}{0.17}
    \begin{tabular}{ccc}
    \includegraphics[width=2.5cm,height=1.6cm]{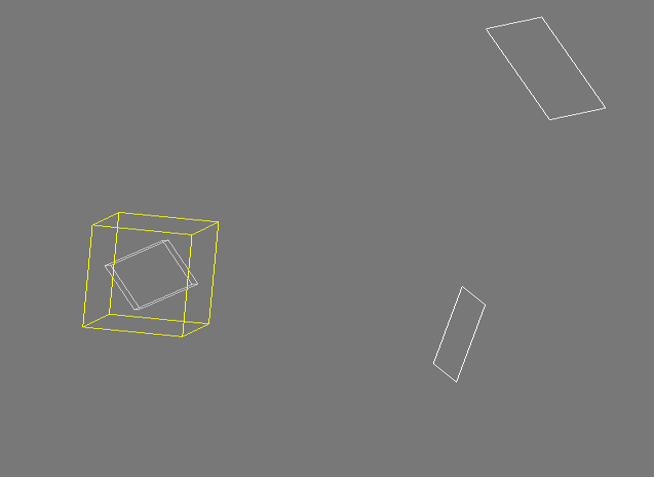}&   \includegraphics[width=2.5cm,height=1.6cm]{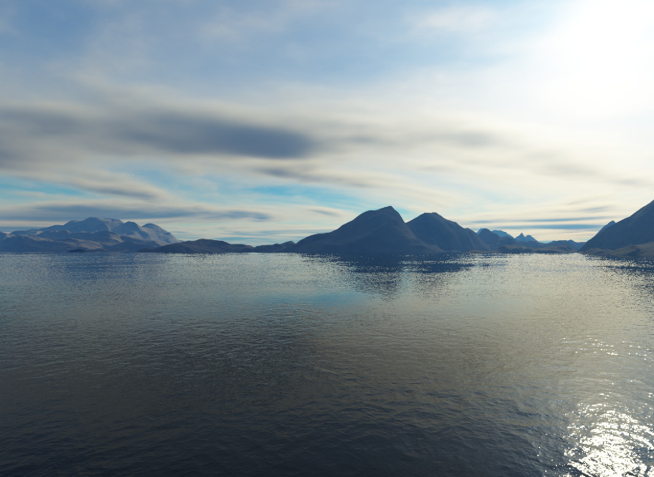}&
    \includegraphics[width=2.5cm,height=1.6cm]{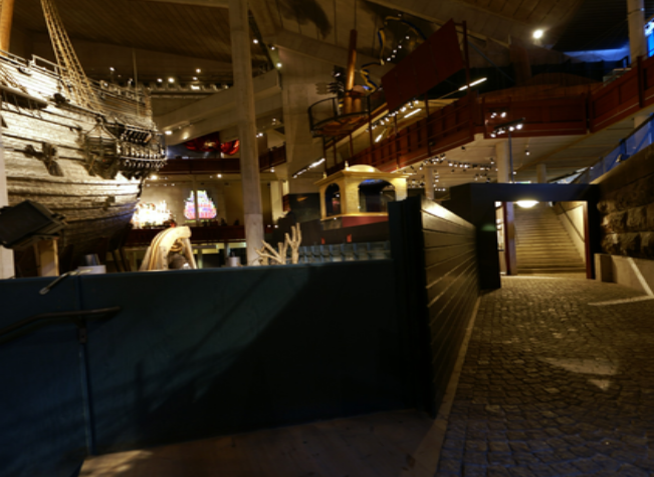}\\

    \includegraphics[width=2.5cm,height=2.3cm]{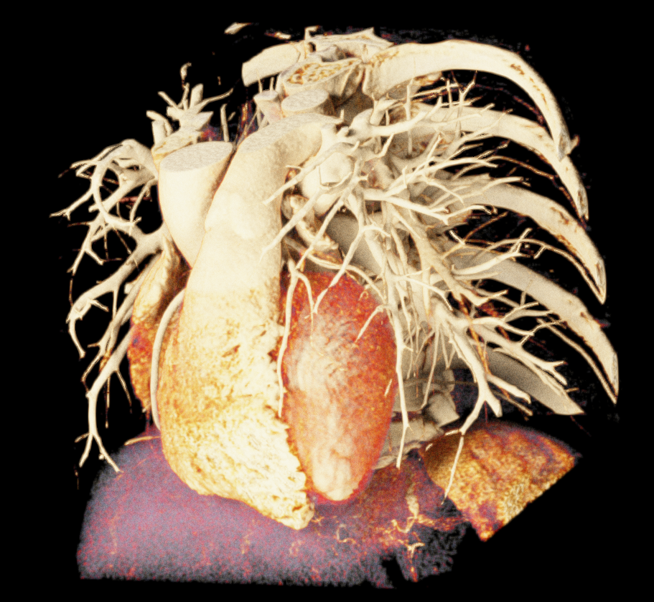}&
    \includegraphics[width=2.5cm,height=2.3cm]{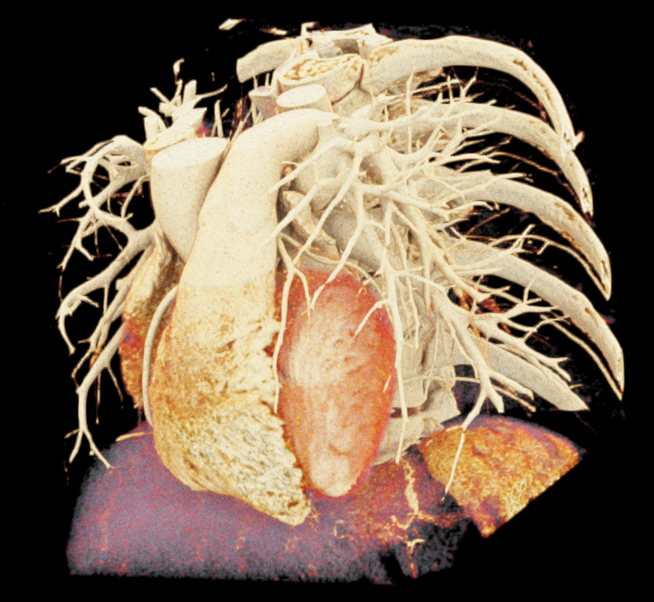}&
    \includegraphics[width=2.5cm,height=2.3cm]{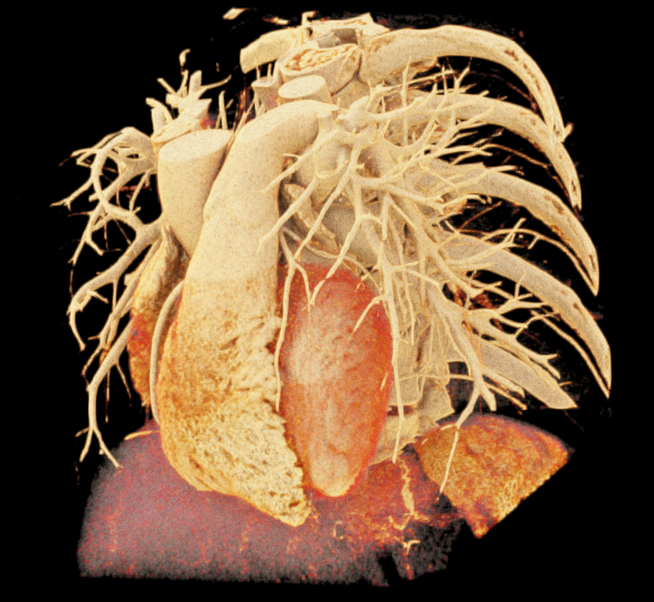}\\
    (a)&(b)&(c)
         
    \end{tabular}
   \caption[figcap]{Cinematic renderings using different lighting strategies: (a) use of two area light sources, (b-c) use of image-based lighting with two different high-dynamic range images; cloudy sky and a Vasamuseum Humus\protect\footnotemark[6]}
   \vspace{-1.5em}
    \label{fig:IBL}
\end{figure}

\textit{Image-based lighting (IBL).} IBL is a computer graphics technique that employs a high-dynamic-range image (HDRI) as a light source. In PBR, background light refers to the light coming from the environment surrounding the rendered scene. We utilized the SoPathTracerBackgroundLight module to support IBL with cubemaps for rendering the volume (Figure \ref{fig:IBL}), demonstrating how different lighting strategies can influence the overall appearance of the rendered image. For instance, the use of two area light sources in the first image provides a more focused and detailed illumination, while IBL with HDRI creates a more realistic and immersive illumination by capturing lighting and reflection information from the surroundings.

\footnotetext[6]{This is the work of Emil Persson, also known as Humus. \url{http://www.humus.name}}

\textbf{Rendering.} Once color, material, and lighting are adjusted, we integrate the SoPathTracer module for rendering. This module utilizes a Monte Carlo path tracing method to simulate light transport through the anatomy and create photo-realistic images. 
Path tracing is a common technique in computer graphics \cite{kajiya1986rendering}, generating paths of scattering events from the camera to light sources, resulting in multi-scattering.
The Monte Carlo integration method solves the following multi-dimensional and non-continuous rendering equation \ref{eq:1}, considering the properties of the scene and the physical interactions of light with the materials: 
\vspace{-1mm}
\begin{equation} \label{eq:1}
L_o(x,\omega_o) = L_e(x,\omega_o) + \int_{\Omega} f_r(x,\omega_i,\omega_o) L_i(x,\omega_i)|\cos\theta| \mathrm{d}\omega_i,
\end{equation}
\vspace{-0.5mm}
\noindent where $L_o(\mathbf{x},\omega_o)$ and $L_e(\mathbf{x},\omega_o)$ are the emitted (from the surface) and outgoing radiances, respectively, at point $\mathbf{x}$ in direction $\omega_o$, $f_r(\mathbf{x},\omega_i,\omega_o)$ is the bidirectional reflectance distribution function (BRDF) that describes how much light is reflected in different directions $\omega_o$ from direction $\omega_i$, $L_i(\mathbf{x},\omega_i)$ is the incoming radiance at point $\mathbf{x}$ from direction $\omega_i$, and $|\cos\theta|$ where theta is the angle between the surface normal and $\omega_i$. 

After setting up all the light sources and material properties, the user can interact with the rendering and adjust the camera projection type. Furthermore, post-processing tools such as the SoPostEffectAmbientOcclusion module were applied to improve the shadowing effect and depth perception. Moreover, the SoVolumeCutting and clip plane modules were added to allow image editing, enabling the exposure and display of specific regions of interest, as well as isolating the heart from adjacent structures such as bones and vessels.

\begin{table*}[h]
\centering
\caption{Likert’s scale evaluation of cinematic renderings across all cardiac structures and characteristics from the commercial and \\open-source solutions.} \label{tab:results}
\resizebox{16 cm}{!}{
\begin{tabular}{|c|cc||cc||cc||cc||cc|}
\hline
\multirow{2}{*}{\textbf{\begin{tabular}[c]{@{}c@{}}Characteristics $\longrightarrow$ \\ Anatomical structure $\downarrow$\end{tabular}}} & \multicolumn{2}{c||}{\textbf{Definition of structure}} & \multicolumn{2}{c||}{\textbf{Depth perception}}         & \multicolumn{2}{c||}{\textbf{Texture appearance}}        & \multicolumn{2}{c||}{\textbf{Fidelity}}                 & \multicolumn{2}{c|}{\textbf{Diagnostic ability}}       \\ \cline{2-11} 
& \multicolumn{1}{c|}{\textbf{Commercial}}  & \textbf{Open} & \multicolumn{1}{c|}{\textbf{Commercial}} & \textbf{Open} & \multicolumn{1}{c|}{\textbf{Commercial}} & \textbf{Open} & \multicolumn{1}{c|}{\textbf{Commercial}} & \textbf{Open} & \multicolumn{1}{c|}{\textbf{Commercial}} & \textbf{Open} \\ \hline \hline
\textbf{Atria}& \multicolumn{1}{c|}{4.00}  & 3.67 & \multicolumn{1}{c|}{4.67} & 4.33& \multicolumn{1}{c|}{3.67} & 3.33  & \multicolumn{1}{c|}{3.67} & 3.33 & \multicolumn{1}{c|}{4.67} & 4.00    \\ \hline
\textbf{Ventricles} & \multicolumn{1}{c|}{4.00} & 3.33 & \multicolumn{1}{c|}{4.67}& 4.33 & \multicolumn{1}{c|}{4.00}  & 3.67 & \multicolumn{1}{c|}{4.00} & 3.67& \multicolumn{1}{c|}{4.67} & 4.00\\ \hline
\textbf{Great Arteries} & \multicolumn{1}{c|}{4.33}  & 4.00  & \multicolumn{1}{c|}{4.67} & 4.33 & \multicolumn{1}{c|}{4.67}& 4.67  & \multicolumn{1}{c|}{4.00}  & 3.67  & \multicolumn{1}{c|}{4.33}  & 4.00 \\ \hline
\textbf{Coronary Arteries}  & \multicolumn{1}{c|}{2.33} & 2.33 & \multicolumn{1}{c|}{4.67}  & 4.33 & \multicolumn{1}{c|}{4.67}  & 4.33   & \multicolumn{1}{c|}{3.67} & 3.33  & \multicolumn{1}{c|}{4.33} & 4.00 \\ \hline \hline
\textbf{Average} & \multicolumn{1}{c|}{\textbf{3.67}}  & \textbf{3.33} & \multicolumn{1}{c|}{\textbf{4.67}} & \textbf{4.33}   & \multicolumn{1}{c|}{\textbf{4.25}}    & \textbf{4.00} & \multicolumn{1}{c|}{\textbf{3.83}} & \textbf{3.50}   & \multicolumn{1}{c|}{\textbf{4.50}}    & \textbf{4.00} \\ \hline
\end{tabular}
}
\vspace{-1.5em}
\end{table*}

\section{Evaluation of cinematic rendering}
\subsection{Assessment protocol}
The overall evaluation consisted of subjective assessments of photo-realistic static snapshots. Three independent domain-expert cardiologists, two cardiac radiologists with 8 years of experience, and one pediatric cardiologist with 15 years of experience, conducted the evaluations. \color{black} The snapshots were acquired in such a way as to enable independent ratings for various anatomical structures, including the atria, ventricles, great arteries, and coronary arteries. A score was required for the snapshots on a Likert scale from 1 to 5 (e.g., very unsatisfied to very satisfied). Specifically, five questions were asked per case, related to the following visual characteristics, adapted from \cite{preim2016survey, steffen2022three}: \vspace{-3mm}
\begin{itemize}[noitemsep]
    \item \emph{Definition of structure} describes the sharpness of the edges, e.g., for performing an anatomical measurement.
    \item \emph{Depth perception} describes the ability to perceive spatial relationships in 3D (e.g., anterior/posterior). 
    \item \emph{Texture appearance} refers to the appearance of the surfaces in terms of their degree of roughness and metalness.
    \item \emph{Fidelity} is a characteristic analyzing the sensation of resembling real cardiac tissue on screen.
    \item \emph{Diagnostic ability} refers to the effectiveness of the rendered images in supporting clinical diagnosis.
\end{itemize}\vspace{-2.15mm}
The questionnaire also included questions about using open-source and commercial solutions for CR. Two questions used a Likert scale to assess the reliability of open-source tools and whether the investigators would recommend them to others. The third question asked for an open-ended opinion about the preference for using open-source or commercial tools for advanced rendering.
\subsection{Results}
Figure \ref{fig:Figure_1} presents the conventional volume alongside cinematic renderings of all five analyzed cases. The top row showcases the volume rendering, while the middle and bottom rows display cinematic renderings created using the commercial solution and open-source framework. In the open-source framework, all cases underwent CR with two area light sources, background white light, and distinct material properties, followed by a post-processing step. Furthermore, noise removal was applied during preprocessing for the congenital cases, as they often had higher noise levels compared to normal adult hearts. Alongside these visual results,
table \ref{tab:results} complements these visual results by providing a comprehensive overview of our evaluation findings. The analysis was performed by taking the average of all anatomical structure scores across all characteristics. All investigators found the renderings from the commercial solution to be superior or equivalent to the open-source alternative reconstructions, for all analyzed anatomical structures and visual characteristics. However, both solutions obtained satisfaction scores of a similar scale, with no significant differences overall. Case 3 notably highlights the importance of CR. Here, the occluded arterial duct (indicated by the red arrow) is challenging to identify using only VR; however, both alternative methods offer comparable visual results. The analysis further reveals minor distinctions in each visual characteristic assessment for the Great Arteries, with variances of approximately 0.33 and 0.34. Interestingly, the texture appearance exhibits remarkable equivalence, emphasizing the effectiveness of both solutions in capturing texture details. Moreover, case 2, involving a ventricular septal defect, exhibited enhanced visual appeal when rendered using an open-source solution; however, the analysis indicates that commercial CRs outperformed across all visual characteristics. 
Furthermore, in perceiving the depth of anatomical structures both CRs performed equally. Additionally, the evaluators found that the edges of anatomical structures were better visualized with the commercial tool, with a mean difference in the definition of structure characteristic of 0.34 with respect to the open-source renderings. We can also observe that the lowest scores for both CRs solutions corresponded to the definition of the coronary arteries, due to their complex anatomy. \color{black}
\section{Discussion and conclusions}
Cinematic rendering for complex cardiac data is a promising 3D visualization tool, which is mainly performed on commercial tools in clinical environments. These tools have high-performance capabilities but are relatively costly, thus not being accessible to everyone. In the present study, we have designed a photo-realistic rendering pipeline using the open-source SDK version of MeVisLab as an alternative.\par
To achieve the best possible cinematic rendering visualizations, we performed a sensitivity analysis to identify the most relevant parameters (e.g., material properties) and their effect on the final results. A limitation of fixing values for material properties is that the same value will not result in the same texture appearance for each case since materials behave differently for each anatomy which also depends on the image quality \cite{elshafei2019comparison}. We also analyzed the effects of lighting and found that the number of light sources and their positions are crucial for creating visually compelling and informative images with higher depth perception. Moreover, area lights offer focused and detailed lighting, while background lighting with HDRI provides more realistic and immersive illumination by capturing lighting and reflection information from the surrounding environment.\par
An evaluation protocol was jointly designed with cardiologists to compare the CRs provided by the developed open-source pipeline and from a commercial solution available in a hospital environment, assessing several visual characteristics of different cardiac structures. Results obtained by three independent cardiologists were consistent in recognizing an overall superior performance of the commercial solution, with slightly higher scores on the Likert scale, especially for the definition of structures, texture appearance, and diagnostic ability. However, the open-source solution had several visual characteristics with a 4.00 or beyond, being similar to the commercial alternative for fidelity and depth perception. Both types of CRs performed worst for the definition of structures due to the low scores given to the coronary arteries. The cardiologists also expressed their satisfaction with the reliability of open-source rendering tools and their recommendation to others for advanced rendering purposes. Their overall preference was for open-source tools, mainly due to their cost-effectiveness. \par
Computationally speaking, both solutions were highly dependent on suitable hardware. The quality of the renderings was dependent on the number of rays to be traced, which in turn depended on the computational power. For example, on a workstation with a GPU RTX 1050, the image was completely rendered in 29 sec (300 iterations) with the open-source solution. However, with a GPU NVIDIA RTX A6000, it only took 3 seconds to completely render the same image. \par
The main strength of the open-source CR pipeline developed in the study is that it allows continuous improvement (\href{https://github.com/upf-gti/cinematic-volume-rendering.git}{GitHub Project}), being a "white box" tool enabling to tune parameters such as transfer functions, materials, and lighting to obtain better visualizations, as per clinician's requirement on specific patients. Considering the accessibility of the open-source solution and the similarity of the corresponding satisfaction scores to the commercial tool, the developed pipeline is an exciting alternative, mainly for educational purposes and to support medical diagnosis as a pre-operative planning tool. Future work will be devoted to a more extensive evaluation with more analyzed cases, evaluators, and used software tools, including web-based solutions democratizing the use of open-source CR in hospitals independently of their resources. 
\acknowledgments{This project has received funding from the European Union’s Horizon 2020 research and innovation programme under grant
agreement No 101016496 (SimCardioTest).}

\bibliographystyle{abbrv-doi}

\bibliography{template}
\end{document}